\documentclass[a4paper]{article}
\usepackage{amsmath}
\usepackage{epsfig,subfigure}
\usepackage{cite}
\date{}
\begin{document}
\title{
\Large{
\textbf{Nonlinear optical properties of a channel waveguide 
produced with crosslinkable ferroelectric liquid crystals
}
}
}
\author{
Valentina S. U. Fazio \footnote{e-mail:
fazio@fy.chalmers.se}, Sven T. Lagerwall \\
\textit{\small{
Department of Microelectronics and Nanoscience,
Liquid Crystal Physics,
}}\\
\textit{\small{
Chalmers University of Technology \&
G\"oteborg University, SE-41296 G\"oteborg, Sweden
}} \\
\\
Vismant Zauls, Sigurd Schrader\\
\textit{\small{
Institute of Physics, University of Potsdam, D-14469 Potsdam, Germany
}}\\
\\
Philippe Busson, Anders Hult\\
\textit{\small{
Department of Polymer Technology, Royal Institute of Technology,
SE-10044, Stockholm, Sweden
}}\\
\\
Hubert Motschmann\\
\textit{\small{
Max-Plank-Institute of Colloids and Interfaces, D-14476 Golm/Potsdam, Germany
}}
}
\normalsize
\normalsize
\maketitle
\begin{abstract}
\noindent
A binary mixture of ferroelectric liquid crystals (FLCs) was
used for the design of a channel waveguide. 
The FLCs possess two important functionalities: a chromophore with a high
hyperpolarizability $\beta$ and photoreactive groups.
The smectic liquid crystal is aligned in layers parallel to the glass plates 
in a sandwich geometry.
This alignment offers several advantages, such as that
moderate electric fields are sufficient to achieve a high degree 
of polar order.
The arrangement was then permanently fixed by photopolymerization
which yielded a polar network possessing a high thermal and 
mechanical stability which did not show any sign of degradation 
within the monitored period of several months. 
The linear and nonlinear optical properties have been measured and all
four independent components of the nonlinear susceptibility
tensor $\overline d$ have been determined. 
The off-resonant $d$-coefficients are remarkably high and comparable 
to those of the best known inorganic materials. 
The alignment led to an inherent channel waveguide for 
p-polarized light without additional preparation steps.
The photopolymerization did not induce scattering sites in the 
waveguide and the normalized losses were less than  2\,dB/cm.  
The material offers a great potential for the design of nonlinear 
optical devices such as frequency doublers of low power laser diodes. 
\end{abstract}
\vspace{10mm}

\noindent
PACS number(s):\\ 
61.30.Gd (orienatational order of liquid crystals; 
electric and magnetic field effects on order) \\
42.65.Ky (harmonic generation, frequency conversion)\\
42.82.Et (waveguides, couplers, and arrays)\\

\vspace{10mm}

\section{Introduction}
Organic materials have a great inherent potential for nonlinear optical (NLO)
devices, such as frequency doublers and a variety of fast switching 
devices.
This potential has been early recognized. 
Meanwhile there is a sound understanding of the correlation between the 
molecular structure and the resulting nonlinear optical properties.
Many NLO chromophores with a remarkably high hyperpolarizability 
$\beta$ have been synthesized.  
At present not the availability of suitable chromophores is the
decisive hurdle for efficient devices, but the fabrication
of suitable structures which have to simultaneously fulfill many
criteria.
Preparation schemes such as the Langmuir-Blodgett
(LB) technique \cite{Petty, Ulman} or poled
polymers\cite{PrasadWilliams} lead to an unwanted intrinsic
dilution of the active moiety within a matrix and in addition these
films usually possess a low thermal or mechanical stability.

The figure of merit of all devices based on second-harmonic effects
is given by the ratio $d_{\text{eff}}^{2} / n^{3}$, where
$d_{\text{eff}}$ is the effective reduced nonlinear optical susceptibility 
and $n$ the refractive index of the material.
The oriented gas model\cite{MotPenArmEnz93} provides a relation 
between molecular and macroscopic quantities.
It states that $d_{\text{eff}}$ is proportional, in first approximation,
to the number density of the NLO chromophores and to the orientational
average, $<\beta>$, of the molecular nonlinear hyperpolarizabilities.
A maximization of $d_{\text{eff}}$ requires a chromophore with a 
large value of $\beta$ arranged in a uniform fashion with a high 
degree of polar order and a high number density.
These demands can be fulfilled by ferroelectric liquid crystals (FLCs).
Sophisticated preparation techniques developed for 
deliberate controlling and manipulating the order in liquid crystal
systems can be utilized for the design of a NLO device.

Liquid crystals (LCs) are ordered organic materials that
intrinsically possess a high number density and a quadrupolar order,
but in general not a dipolar one\cite{deGennes}.
In chiral smectic C (SmC$^{*}$) liquid crystals, or
FLCs, however, the molecular symmetry allows a local dipolar order
perpendicular to the director\cite{MeyLieStrKel75} which can be extended to the
whole sample either by orienting all the molecular dipoles
in an external electric field or by surface constraints like the one 
used in surface-stabilized FLCs (SSFLCs) \cite{ClaLag80} which are 
characterized by a macroscopic polarization in the field-free state 
and in which the director can be switched between two stable states 
(bistable).
In absence of a sufficient external field, the molecules are arranged in a 
helical fashion proper of the chiral phase, which has $D_{\infty}$ 
symmetry and which does not posses a macroscopic polarization.

A lot of effort has been invested in the last decade in the 
synthesis and characterization of FLCs
\cite{WalRosClaSha91, WalRosSieReg91, WalDyeCobSie96, LiuRobJohWal91}
and FLC polymers \cite{KapZenTwiNgu90, WisZenRedMon94, NacRatBarKel95, 
SveHelSkaAnd98} for second-harmonic generation (SHG).
Recently, bent-shaped molecules with very high nonlinear optical
hyperpolarizabilities have been synthesized\cite{MacKenWarHep98}.
Despite all these accomplishments, the low thermal and mechanical stability
of these systems is the reason why they did not gain a practical
level of relevance.
This issue is addressed in the present publication.

The FLCs used in our study possess a photoreactive group and can
be crosslinked to give thermally and mechanically stable
polar polymer networks\cite{HermannDavid97, TroOrrSahGed96, HulSahTroLag96,
TroSahGedHul96}.
The material can first be aligned by an electric field and then the 
polar order can be \lq\lq frozen-in\rq\rq by photopolymerization.
The macroscopic polarization is now an intrinsic property of the polymer
film which is neither ferroelectric nor truly liquid crystalline,
but rather a \textit{pyroelectric polymer} (PP), characterized by a
macroscopic polarization which cannot be switched\cite{HulSahTroLag96}.

SHG in similar polymers has been studied in planar cells
\cite{TroOrrSahGed96, HulSahTroLag96, TroSahGedHul96,
LinHerOrtArn98} where the glass plates were treated with
unidirectionally rubbed polyimide for planar alignment
(smectic layers standing perpendicular to the plates).
Usually refractive indices and thickness of these aligning layers, such as
polyimide, are unknown resulting in complications in the data
interpretation.
Moreover, because of the surface constraint, high electric fields
are required to unwind the smectic helix.

Instead, we use quasi-homeotropic orientation.
In this geometry all the (four) independent reduced second-order nonlinear
susceptibilities can be determined, which is not the case in the planar
geometry (see for instance \cite{HermannDavid97}).
There are a number of other advantages in this geometry as well.
First of all, a very high quality of the alignment is obtained almost 
automatically when the layers are parallel to the glass plates and no 
aligning precoating is necessary.
Second, there is essentially no threshold for the azimuthal motion of 
the director which continuously follows the electric field.
An eventual small threshold could only result from surface pinning 
effects.
Moreover, even the threshold for unwinding the smectic helix 
generally is lower in the quasi-homeotropic case than in the planar 
(bookshelf) one.
Finally, in the homeotropic geometry the sample is a channel waveguide
for p-polarized light without any additional preparation steps.

\section{Material}
\begin{figure}
\begin{center}
\epsfig{file=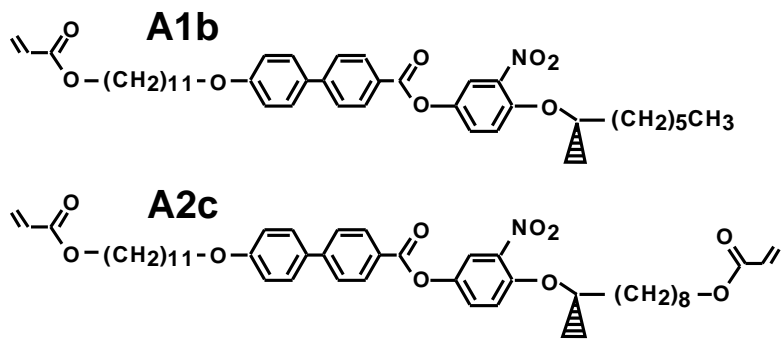, width=0.85\textwidth}
\caption{
\label{uno}
\small{Ferroelectric liquid crystal monomers used in this experiment.}
}
\end{center}
\end{figure}

The chemical structures of the FLC monomers used in this work
are shown in Figure \ref{uno}.
The monomer \textsf{A2c} is terminated by two photoreactive end-groups
that allow the formation of a stable polymeric network by
photopolymerization\cite{TroOrrSahGed96}.
Within the polymer network the polar order of the monomeric system is 
permanently fixed (\lq\lq frozen\rq\rq).
The monomeric FLC forms a  highly twisted  SmC$^{*}$
phase with a pitch of 0.4\,$\mu$m\cite{HerRudLagKom98} and therefore
a fairly  high DC electric field (more than 3\,V\,$\mu$m$^{-1}$) is 
required to unwind the helix.
The monomer \textsf{A1b} possesses only a single
photoreactive end-group and only side-chain polymers are obtained after
photopolymerization, but its helix can be unwound by a DC electric field of a
few hundred mV\,$\mu$m$^{-1}$ in planar cells \cite{HermannDavid97}.
Both monomers possess a quite large spontaneous polarization
(about 190\,nC\,cm$^{-2}$ at room temperature) \cite{HerRudLagKom98}.

In this work a mixture of 60\,{\%} \textsf{A1b} and 40\,{\%}
\textsf{A2c} was used which combines the desired
features of both monomeric systems: the low voltage to unwind the 
helix and establish a homogeneously polarised state and the possibility 
to freeze it by formation of a stable polymer network.
The photopolymerization was enabled by adding a very small quantity
(less than 1\,{\%} in weight) of the photoinitator Lucirin TPO (BASF) 
to the mixture.
The phase behavior of the individual monomers and of the mixture
is listed in Table \ref{table_uno}.
\begin{table}
\begin{center}
\begin{tabular}{cccc}
\hline
compound & heating & cooling \\
\hline
\textbf{A1b} & K 29 SmC$^{*}$ 58 SmA$^{*}$ 71 I &
I 62 SmA$^{*}$ 50 SmC$^{*}$ 0 K & \protect\cite{HermannDavid97}\\
\textbf{A2c} & K 29 SmC$^{*}$ 33 SmA$^{*}$ 39 I &
I 34 SmA$^{*}$ 29 SmC$^{*}$ 8 K & \protect\cite{TroOrrSahGed96}\\
\textbf{A1b/A2c 60/40} & & I 44 SmA$^{*}$ 31 SmC$^{*}$ 9 K & this work\\
\hline
\end{tabular}
\caption{
\label{table_uno}
\small{Thermal transitions for the two ferroelectric liquid crystal
monomers and the mixture used in this work (temperature in degrees Celcius;
K = crystal, I = isotropic).
}
}
\end{center}
\end{table}
%

%
%
\section{Sample preparation}
\begin{figure}
\begin{center}
\epsfig{file=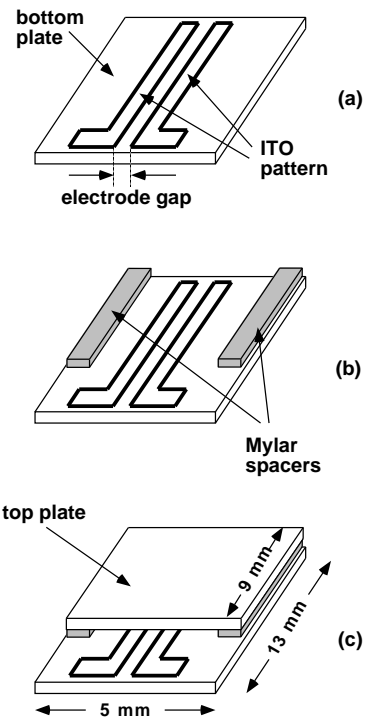, width=0.40\textwidth}
\caption{
\label{due}
\small{Manufacturing steps.
(a) Patterning of the ITO electrodes. The gap between the
electrodes is 100\,$\mu$m wide. The thickness of the ITO layer is
about 15\,nm.
(b) Deposition of Mylar spacers of 13\,$\mu$m thickness.
(c) The top glass plate is glued on, to complete the sandwich cell.}
}
\end{center}
\end{figure}

The geometry of the cell is shown in Figure \ref{due}.
The bottom plate is equipped with two parallel ITO electrodes
with 100\,$\mu$m gap patterned by photolithography.
No special surface treatments were used.
The two plates were spaced with polyester films (Mylar) of various 
thickness.

\begin{figure}
\begin{center}
\epsfig{file=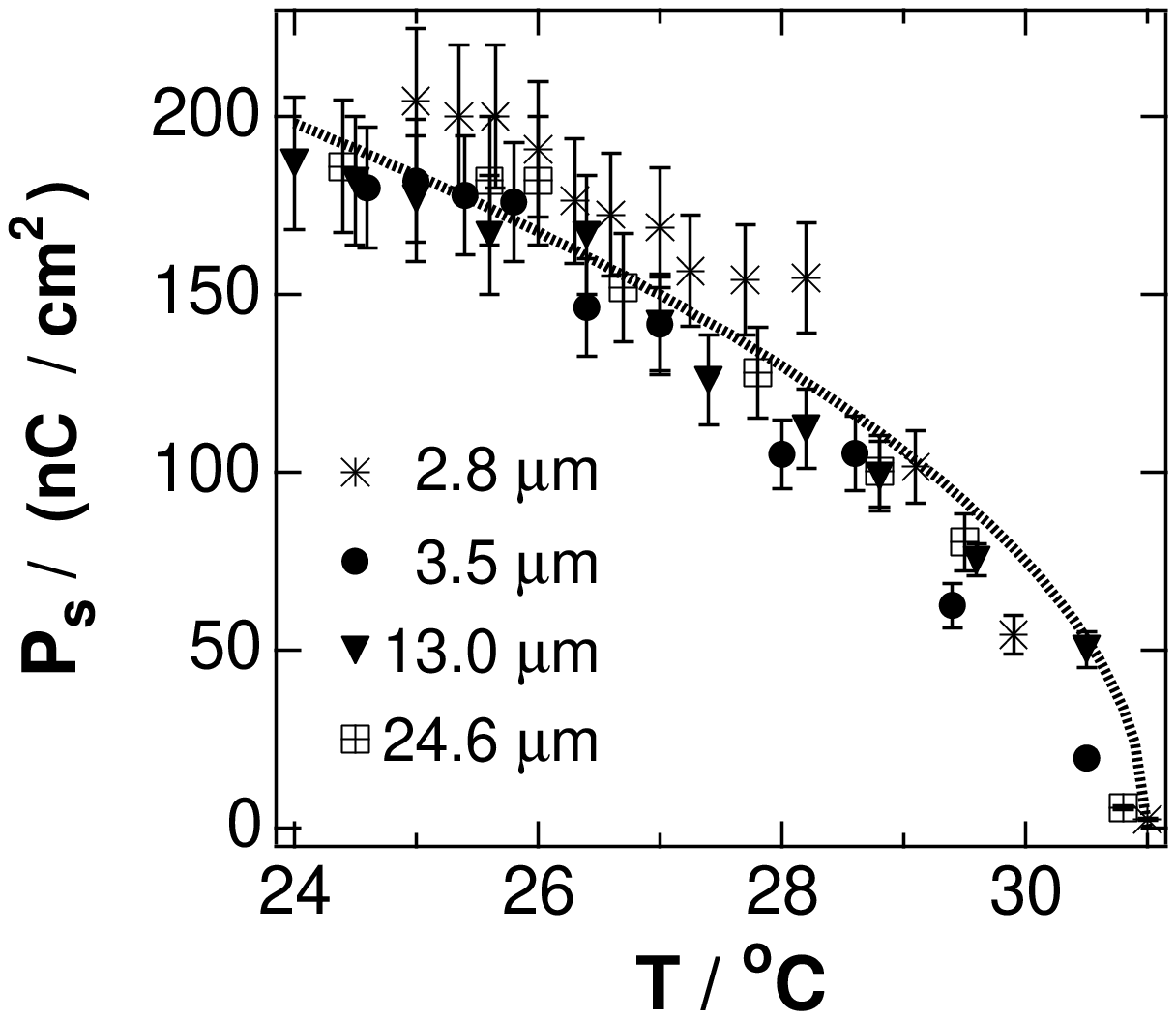,width=0.65\textwidth}
\caption{
\label{Ps}
\small{Spontaneous polarization as a function of temperature for the
mixture used in this experiment.
The markers represent $P_{s}$ values measured in four cells of 
different thickness, the dashed line is a fit to the function
$P_{s} = A \sqrt{B - T}$\cite{deGennes}, where $A$ and $B$ are fitting parameters
($A=74.9\,\text{nC cm}^{-2} {}^{\circ}{\text{C}}^{-\frac{1}{2}}, B =
31\,{}^{\circ}$C), and $T$ is the temperature in degrees Celcius.}
}
\end{center}
\end{figure}

The cells were inserted into a polarising microscope equipped with a 
hot stage and filled with the FLC mixture in the isotropic phase.
During cooling to room temperature a small low-frequency AC electric field
(0.1-0.2\,V\,$\mu$m$^{-1}$) was applied across the electrodes,
both to facilitate the desired smectic order and to measure
the electrooptic response.
By integrating the transient current on field reversal the spontaneous 
polarization $P_{s}$  is obtained.
The temperature dependence of $P_{s}$ is shown in Figure \ref{Ps}.
At room temperature a value of $\approx$ (190 $\pm$ 20)\,nC\,cm$^{-2}$ 
was measured, which is in agreement with the values found for the single 
monomers\cite{HerRudLagKom98}.

\begin{figure}
\begin{center}
\epsfig{file=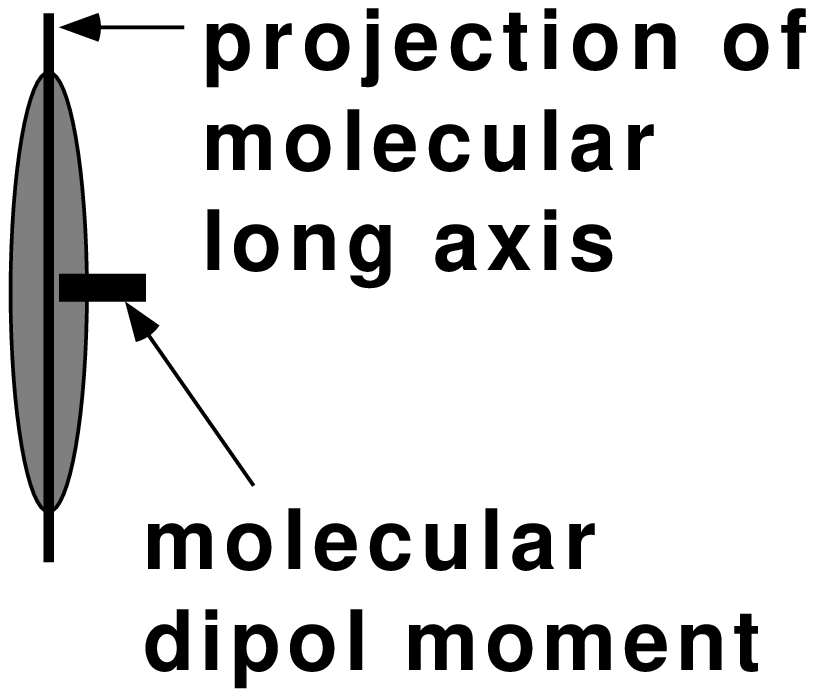,width=0.20\textwidth}
\hspace{10mm}
\epsfig{file=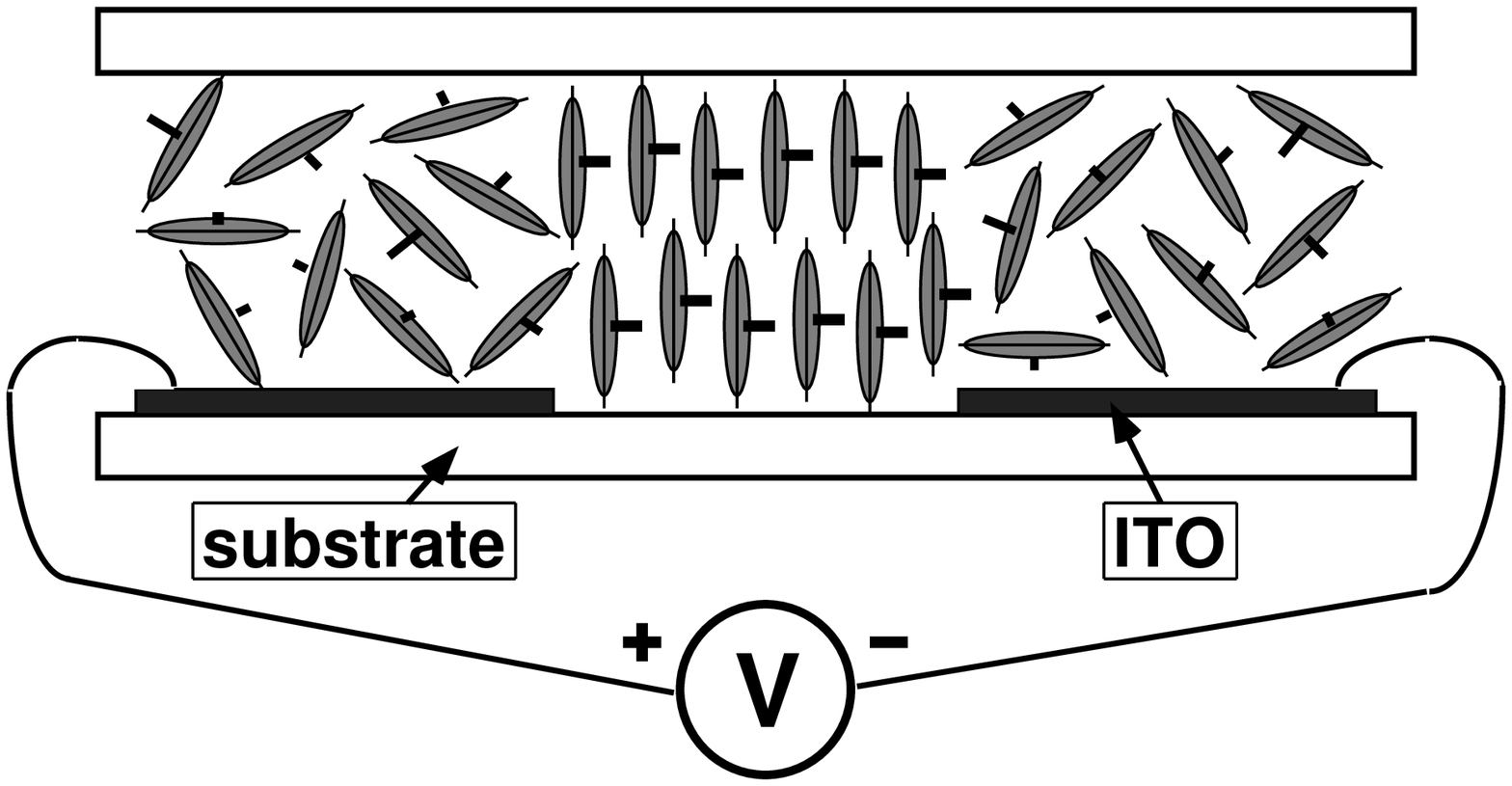,width=0.55\textwidth}
\caption{
\label{cross_sec}
\small{Cross section of a cell.
In the gap between the electrodes the FLC material is oriented 
by the AC/DC field.
The dipole moments are aligned along the cell plates perpendicular to
the electodes stripes.
The smectic tilt plane is perpendicular to the page.
In the channel between the electrodes we have a microscopic polarization 
whereas outside it we have a disordered non-polar state (the liquid 
crystalline disorder in these regions is somewhat exaggerated in the 
figure).
}
}
\end{center}
\end{figure}

\begin{figure}
\begin{center}
\epsfig{file=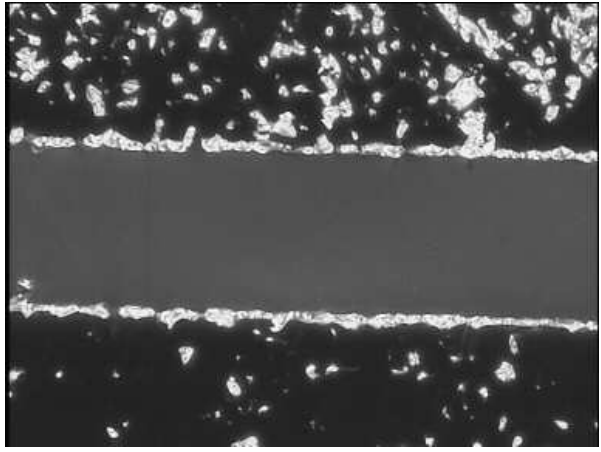,width=0.65\textwidth}
\caption{
\label{foto}
\small{Micrograph of a cell after polymerization.
The PP molecules are uniformly oriented in the channel between
the electrodes.
The cell thickness is 13\,$\mu$m and the electrode gap is 100\,$\mu$m. }
}
\end{center}
\end{figure}

At room temperature a DC electric field ($\approx$ 50\,mV\,$\mu$m$^{-1}$)
was applied to unwind the smectic helix and orient the FLC
molecular dipole moments in the electrode gap (see Figure \ref{cross_sec}).
Since the gap between the electrodes is larger than the cell size by an 
order of magnitude, the electric field inside the channel is approximatly 
horizontal and edge effects can be neglected.
As a result a macroscopic domain of uniform orientation was formed 
between the electrodes.
The samples were then irradiated with UV light.
The photopolymerization led to the formation of a stable network:
thereafter a temperature up to 160\,${}^{\circ}$C did not induce any
phase transition and the polar order could not be reversed by an 
electric field.
In other words, the polar order induced by the external field had 
been frozen.
The photopolymerization did not induce any defects visible by optical
means and the alignment was uniform within the stripe.
Figure \ref{foto} is a micrograph of a cell after photopolymerisation.
The stripe appeared dark gray between crossed polarisers. 
Rotation of the sample between crossed polarizers did 
not essencially change any feature.
This suggests that in such thick cell the smectic helix is not 
completely unwound and that the optic axis as well as the polarization is
not completely homogeneous across the sample.
Yet, we can consider the PP essentially homeotropically aligned.

\section{Second-harmonic generation experiment}
\begin{figure}
\begin{center}
\epsfig{file=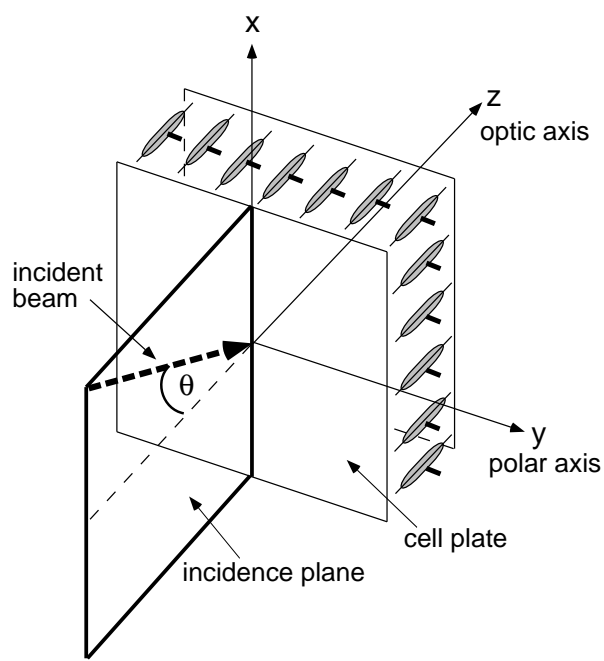,width=0.65\textwidth}
\caption{
\label{HOM_incidence}
\small{Geometry and definition of the coordinate system of the SHG
experiment in the channel region.
The macroscopic dipole moment of the polymer network is oriented
along the $y$-axis, parallel to the glass plates.
Allthoug the director is not perfecly homogeneous across the cell, the
effective optic axis is along $z$, perpendicular to the glass plates.
$\theta$ is the internal incidence angle.}
}
\end{center}
\end{figure}

SHG requires noncentrosymmetry and thus can only occur within the
channel region of our waveguide.
Figure \ref{HOM_incidence} shows the geometry and definition of the
frame of reference for modelling SHG.

\begin{figure}
\begin{center}
\epsfig{file=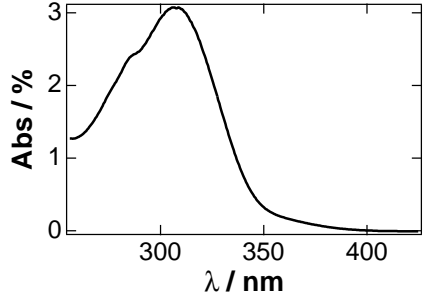,width=0.55\textwidth}
\caption{
\label{abs_spectrum}
\small{Adsorption spectrum of the mixture \textsf{A1b}/\textsf{A2c}
60/40 in chloroform solution 0.13\,mM.
The material is transparent at the wavelengths of the fundamental
(1064\,nm) and the second harmonic (532\,nm) light.}
}
\end{center}
\end{figure}

\begin{figure}
\begin{center}
\epsfig{file=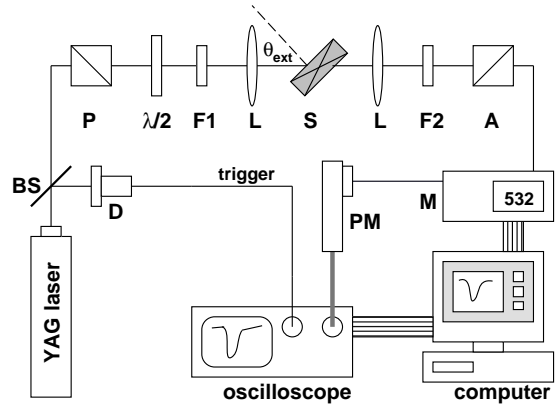,width=0.6\textwidth}
\caption{
\label{SHGsetup}
\small{Setup for second-harmonic generation.
\textsf{P}: polarizer; \textsf{F1}: visible-cut filter; \textsf{L}:
lens; \textsf{S}: sample; $\theta_{\mbox{ext}}$: external incidence
angle; \textsf{F2} fundamental-cut filter;
\textsf{A}: analyzer; \textsf{BS}: beam splitter; \textsf{D}: detector;
\textsf{M}: monochromator; \textsf{PM}: photomultiplier tube.}
}
\end{center}
\end{figure}

The unwound SmC$^{*}$ liquid crystal phase belongs to the
C$_{2}$ symmetry group.
The second-order reduced susceptibility tensor, $\overline d$,  of
this group contains four independent coefficients\cite{PrasadWilliams}:
\begin{equation}
\overline d =
\begin{pmatrix}
0 & 0 & 0 & d_{14} & 0 & d_{16} \\
d_{16} & d_{22} & d_{23} & 0 & d_{14} & 0 \\
0 & 0 & 0 & d_{23} & 0 & d_{14} \\
\end{pmatrix}.
\label{dmatrix}
\end{equation}
In the homeotropic geometry only s-to-s or p-to-s conversions are allowed
\cite{HermannDavid97}.
The effective susceptibilities which govern the conversions can be 
expressed in terms of the components of the $\overline d$ tensor as:
\begin{align}
\mbox{s-to-s} \,\,\,\,\,\,\,\,\,\,
d_{\text{eff}} &= d_{22}, \label{dss}
\\
\mbox{p-to-s} \,\,\,\,\,\,\,\,\,\,
d_{\text{eff}} &= d_{16} \cos^{2} \theta + d_{23} \sin^{2} \theta + 2 d_{14}
\sin \theta \cos \theta, \label{dps}
\end{align}
where $\theta$ is the fundamental light incidence angle at the 
glass/polymer interface (internal incidence angle, Figure \ref{HOM_incidence}) 
and $\theta_{\omega} = \theta_{2 \omega} = \theta$.
Eqn. \ref{dss} and \ref{dps} contain all the four independent
coefficients of the $\overline d$-tensor.
Hence, in homeotropic geometry, all four independent second-order reduced
susceptibilities can be determined by measuring $d_{\text{eff}}$ as a 
function of the incidence angle $\theta$.
This measurenet was performed in transmission with a YaG laser 
(1064\,nm, 35\,ps, 10\,Hz) in 13\,$\mu$m thick cells.
The material is transparent in the range of wavelengths of interest
as can be concluded from the absorption spectrum 
shown in Figure \ref{abs_spectrum}.
Therefore, our measurements yielded the off-resonant values of $d_{\text{eff}}$.
A sketch of the SHG experiment is shown in Figure \ref{SHGsetup}.
The samples were rotated about the polar axis ($y$-axis in Figure
\ref{HOM_incidence}) and the second-harmonic light intensity was
recorded versus the rotation angle $\theta_{\text{ext}}$.
Due to the small size of the channel region (only 100\,$\mu$m wide)
the alignment is rather critical and we had to ensure that the 
goniometer axis of rotation coincides with the polar axis of the PP 
cell.

\begin{figure}
\begin{center}
\epsfig{file=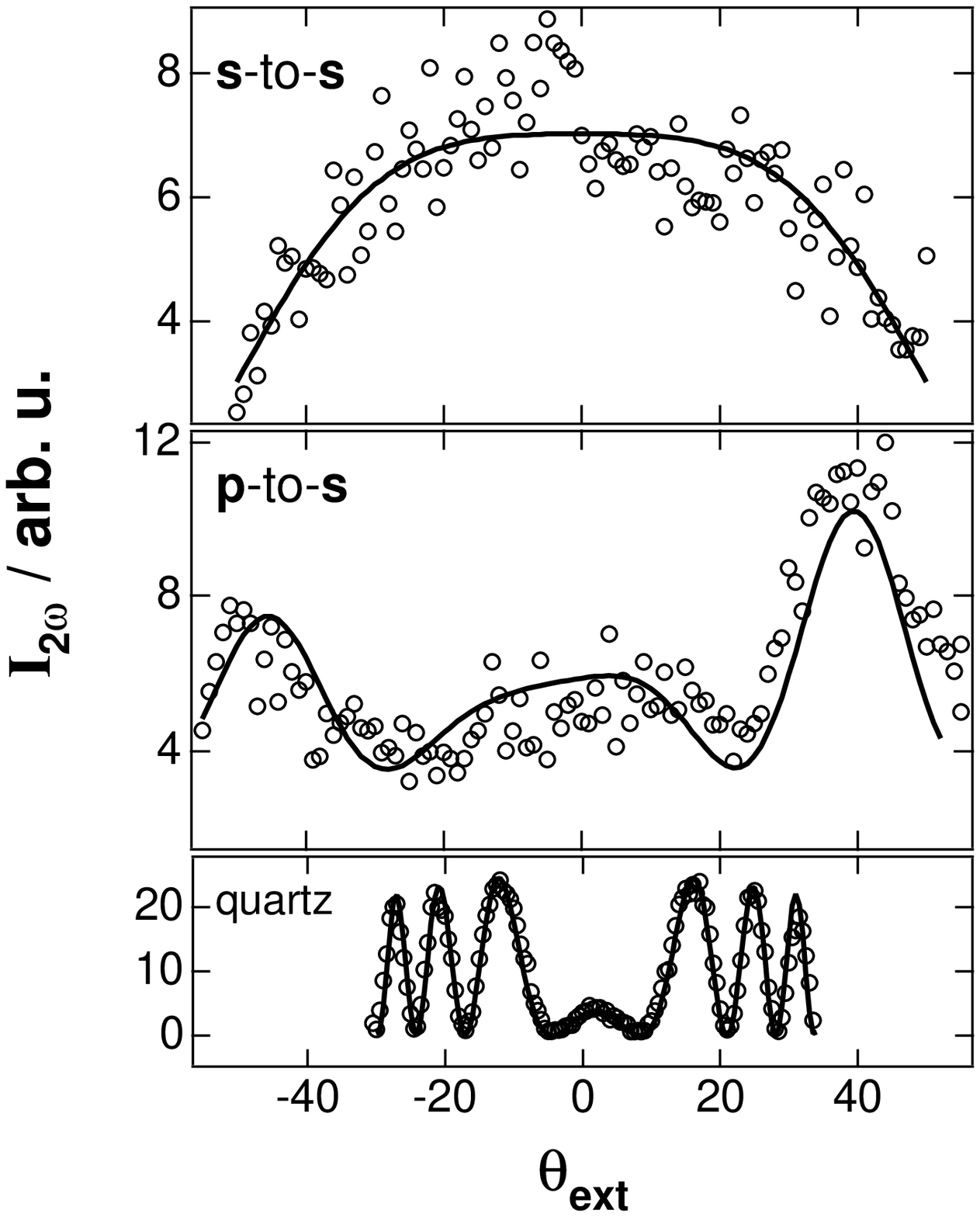,width=0.75\textwidth}
\caption{
\label{SHG_res}
\small{Second-harmonic light intensity versus external angle of incidence
for s-to-s and p-to-s conversions, and for the quartz crystal of
reference.
The dots are the experimental values and the lines correspond to fits
according to\protect\cite{JerKur70}, with the following parameters:
glass plate refractive indices, $n_{\text{g}}(\omega) = 1.46$ and
$n_{\text{g}}(2 \omega) = 1.50$;
FLC ordinary refractive indices, $n_{\text{o}}(\omega) = 1.47$ and
$n_{\text{o}}(2 \omega) = 1.52$ \protect\cite{LinHerOrtArn98};
FLC extraordinary refractive indices, $n_{\text{e}}(\omega) = 1.57$ and
$n_{\text{e}}(2 \omega) = 1.62$ \protect\cite{LinHerOrtArn98};
cell thickness 13\,$\mu$m;
second-order susceptibility of the quartz used for calibration,
$d_{11} = (0.32 \pm 0.04$)\,pm\,V$^{-1}$\protect\cite{Weber}.
The values of the second-order reduced susceptibilities calculated
from the fits are listed in Table \protect\ref{d_coeff}.
}
}
\end{center}
\end{figure}

The SHG intensities as function of the external angle of incidence
are shown in Figure \ref{SHG_res} for both s-to-s and p-to-s polarizations. 
The dots are the experimental values and the lines are fits
\cite{JerKur70} that yielded the values of the four independent 
$d$-coefficients listed in Table \ref{d_coeff}.
\begin{table}
\begin{center}
\begin{tabular}{ccc}
\hline
$d_{22}$ & = & (0.76 $\pm$ 0.10)\,pm\,V$^{-1}$ \\
$d_{16}$ & = & (0.63 $\pm$ 0.09)\,pm\,V$^{-1}$ \\
$d_{23}$ & = & (1.26 $\pm$ 0.16)\,pm\,V$^{-1}$ \\
$d_{14}$ & = & (0.13 $\pm$ 0.02)\,pm\,V$^{-1}$ \\
\hline
\end{tabular}
\caption{
\label{d_coeff}
\small{The four independent $d$-coefficients as calculated from
the fits in Figure \protect\ref{SHG_res}.}
}
\end{center}
\end{table}
The values are comparable to those of quartz due to the large molecular
hyperpolarizabilities combined with the high degree of orientational order 
and the high number density of the active chromophores in our system.
Some of the coefficients exceed the values measured in planar geometry
by 30-50 {\%}\cite{LinHerOrtArn98}, which shows the superior
features offered by the homeotropic alignment.

\section{Waveguide characteristics}
The structure is a channel waveguide for p-polarised light (TM modes).
Assuming homeotropic alignment between the electrodes, the propagation 
characteristics of TM modes in the channel region are dominated by 
the extraordinary refractive index of the PP, $n_{\text{e}}$. 
On the other hands, outside the channel region p-polarised light will
see a refractive index which is somewhere in between $n_{\text{e}}$ and 
$n_{\text{o}}$.
Since $n_{\text{e}} > n_{\text{o}}$ at all wavelengths
(the values $n_{\text{e}}$ and $n_{\text{o}}$ are reported in the 
caption of Figure \ref{SHG_res}), TM modes are three-dimensionally
confined in the channel region and no additional preparation processes are
required to obtain this desired feature.
Here again the homeotropic geometry is advantageous.
The corresponding arrangement in planar cells, where the alignment
is uniform in the whole cell, leads only to slab waveguides.

The losses for p-polarised light at 1064\,nm were measured by monitoring
the output power at different lengths of the guide.
\textit{End-fire} coupling was used to excite waveguide modes.
The transmitted power was first measured with the 9\,mm long
waveguide. 
Then the guide was shortened to 4\,mm and the transmitted power was
measured again using the very same arrangement.
The loss coefficient was found to be
\begin{equation}
\alpha\, = \, (1.07 \pm 0.95 )\, \, \text{dB} \,\, \text{cm}^{-1}.
\label{alphaTOT}
\end{equation}
The upper limit of the losses is on the order of 2\,dB/cm which meets already
the demands imposed by nonlinear optical devices, as for instance frequency
doublers in waveguide format, and demonstrates the inherent potential of
FLCs.
Most devices require only an interaction length of 1--2\,mm.
It is also remarkable that the scattering characteristics did not change with
the polymerization even though this step is always accompanied by changes in the
molecular arrangement.
It is known for instance from LB films that polymerization
induces scattering sites as the result of the formation of additional
bonds.
Apparently the high fluidity of the monomeric LC system is capable
to heal these types of distortions.

There are two kinds of scattering losses in optical
waveguides\cite{Hunsperger}: volume and surface scattering losses.
Volume scattering is caused by defects in the waveguide volume
(scattering centers) and depends on the number of defects and on their
relative dimension to the wavelength of the light, $\lambda$.
If the defects are small compared to $\lambda$, the losses due to
volume scattering will be negligible compared to the ones due to
surface scattering.
This is the case in our samples: no defects were observed in the channel 
region either before or after polymerization, and no scattering
of light due to the polymerization was observed.

Surface scattering losses occur at each reflection of the traveling 
wave in the waveguide\cite{Hunsperger}.
Even if the surfaces are very smooth this kind of losses can be
significant.
The surface scattering loss coefficient in a symmetric waveguide in
the approximation of well-confined modes is\cite{Tie71}:
\begin{equation}
\alpha_{\text{s}} = A^{2}   \frac{1}{2} \frac{\cos^{3}
(\frac{\pi}{2} - \theta_{m})}
{\sin(\frac{\pi}{2} - \theta_{m})}  \frac{1}{t},
\label{alphaS}
\end{equation}
where
\begin{equation}
A = \sqrt{2} \frac{4 \pi n_{m}}{\lambda} \sigma_{\text{gf}}.
\label{A}
\end{equation}
$\theta_{m}$ and $n_{m}$ are the internal reflection angle
and the refractive index of the \textit{m}-th guided mode, respectively;
$t$ is the waveguide thickness;  $\lambda$ is the vacuum wavelength;
$\sigma_{\text{gf}}$ is the variance of the surface roughness at the
glass/film interface.

\begin{center}
\begin{figure}
\epsfig{file=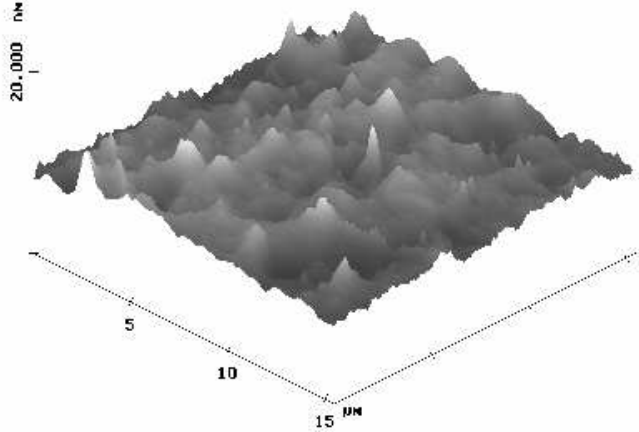, width=0.7\textwidth, angle=-90}
\caption{
\label{afm}
AFM scan of the polymer surface in the channel region.
The variance of the surface roughness is
$\sigma_{\text{gf}}$ = 2.3\,nm $\pm$ 1.0\,nm.
}
\end{figure}
\end{center}

To measure the roughness of the polymer film, a hydrophobic quartz 
glass plate was used as top plate for a 13\,$\mu$m thick cell. 
Being hydrophobic, the plat could be removed without damaging 
the PP\cite{note2}.
The polymer surface was then studied with atomic force microscopy.
Several scans were made and the variance of the surface roughness
in the channel region was measured, $\sigma_{\text{gf}}$ = 2.3\,nm
$\pm$ 1.0\,nm.
A representative AFM picture is shown in Figure \ref{afm}.
The surface appears extremely smooth, and the roughness is almost
entirely due to that of the quartz glass plate (which was previously 
measured with an Alpha-step).
Taking into account all the TM modes supported by the 13\,$\mu$m thick
waveguide and the variance of the surface roughness just measured,
the total surface loss coefficient could be calculated:
\begin{equation}
\alpha_{\text{s,TOT}}= (0.42 \pm 0.23)\,\, \text{dB} \,\, \text{cm}^{-1}.
\label{alphaSCA}
\end{equation}
This means that the optical losses are mainly due to
the coupling of the light in and out of the waveguide.

\section{Conclusions}
The fabrication of a channel waveguide on the basis of ferroelectric
liquid crystals has been demonstrated.
The FLC monomers can be quasi-homeotropically aligned by an external
electric field, resulting in a high degree of order and number density.
This order can be frozen-in by photopolymerization which yields a network
with a high mechanical and thermal stability.
The measured losses of the waveguide are about 1--2\,dB/cm
(normalized to the fraction of the power guided in the organic material)
and already meet the requirements imposed by nonlinear optical devices.
The network formation does not lead to a degradation of the quality of the
waveguide and the fluidity in the LC system heals distortions caused by the
formation of new chemical bonds.
All independent nonlinear optical coefficients have been determined and
the high number density of the chromophores and their high degree of
orientational order lead to remarkably high $d$ values.

The use of photopolymerizable FLCs offers rather appealing
advantages compared to the established preparation techniques such as
Langmuir-Blodgett films or poled polymers.
Due to intrinsic peculiarities of these techniques the number density
remains rather low with only moderate levels of orientational order.
The FLC system does not suffer from dilution of the chromophores and 
exhibits higher degree of orientational order.
The preparation techniques to manipulate order in LC systems are fairly 
sophisticated and allow the design of nonlinear optical devices such as 
frequency doublers of low power diodes or opto-opto switches based on 
cascading $\chi^{(2)}$ interactions.
Critical issues, as for instance the problem of phase-matching or the problem
arising from a small overlap integral in waveguide structures, have to be
duly considered in this new class of materials.
With the possibility of locally controlling the polar order, it seems 
that FLCs have a great potential in this area.

\section{Acknowledgments}
Dr. T. Henning from the Swedish Nanometer Laboratory is greatfully
acknowledged for the production of the photolithography masks used to make
the electrode patterns.
The authors are grateful to Dr. S. Leporatti and M. Schneider for the 
AFM images.
Valentina S. U. Fazio and S. T. Lagerwall acknowledge the TMR 
European programme (contract number ERBFMNICT983023) and the
Swedish Foundation for Strategic Research for financial support.
P. Busson acknowledges the financial support from the Swedish Research 
Council for Engineering Science (TFR, grant 95-807).

\bibliography{journal2,nlo21}

\begin{thebibliography}{10}

\bibitem{Petty}
M.~C. {Petty}.
\newblock {\em {Langmuir}-{Blodgett} films: an introduction}.
\newblock {Cambridge} {University} {Press}, 1996.

\bibitem{Ulman}
A.~{Ulman}.
\newblock {\em An introduction to ultrathin organic films: from
  {Langmuir}-{Blodgett} to self-assembly}.
\newblock {Academic} {Press} {Boston}, 1991.

\bibitem{PrasadWilliams}
P.~N. {Prasad} and D.~J. {Williams}.
\newblock {\em Introduction to nonlinear optical effects in molecules and
  polymers}.
\newblock {John} {Wiley} \& {Sons}, 1991.

\bibitem{MotPenArmEnz93}
H.~{Motschmann}, T.~{Penner}, N.~{Armstrong}, and M.~{Enzenyilimba}.
\newblock {\em J. Phys. Chem}, 97:3933, 1993.

\bibitem{deGennes}
P.~G. de~{Gennes}.
\newblock {\em The physics of liquid crystals}.
\newblock {Oxford} {University} {Press}, 1974.

\bibitem{MeyLieStrKel75}
R.~B. {Meyer}, L.~{Liebert}, L.~{Strzeleki}, and P.~{Keller}.
\newblock {\em J. Physique}, 36:L69, 1975.

\bibitem{ClaLag80}
N.~{Clark} and S.~T. {Lagerwall}.
\newblock {\em Appl. Phys. Lett.}, 36:899, 1980.

\bibitem{WalRosClaSha91}
D.~M. {Walba}, M.~{Blanca Ros}, N.~A. {Clark}, R.~{Shao}, K.~M. {Johnson},
  M.~G. {Robinson}, J.~Y. {Liu}, and D.~{Doroski}.
\newblock {\em Mol. Crys. Liq. Crys.}, 198:51, 1991.

\bibitem{WalRosSieReg91}
D.~M. {Walba}, M.~{Blanca Ros}, T.~{Sierra}, J.~A. {Rego}, N.~A. {Clark},
  R.~{Shao}, M.~D. {Wand}, R.~T. {Vohra}, K.~E. {Arnett}, and S.~P. {Velsco}.
\newblock {\em Ferroel.}, 121:347, 1991.

\bibitem{WalDyeCobSie96}
D.~M. {Walba}, D.~J. {Dyer}, P.~L. {Cobben}, T.~{Sierra}, J.~A. {Rego}, C.~A.
  {Liberko}, R.~{Shao}, and N.~A. {Clark}.
\newblock {\em Ferroel.}, 179:211, 1996.

\bibitem{LiuRobJohWal91}
J.-Y. {Liu}, M.~G. {Robinson}, K.~M. {Johnson}, D.~M. {Walba}, M.~{Blanca Ros},
  N.~A. {Clark}, R.~{Shao}, and D.~{Doroski}.
\newblock {\em J. Appl. Phys.}, 70(7):3426, 1991.

\bibitem{KapZenTwiNgu90}
H.~{Kapitza}, R.~{Zentel}, R.~J. {Twieg}, C.~{Nguyen}, S.~U. {Vallerien},
  F.~{Kremer}, and C.~G. {Wilson}.
\newblock {\em Adv. Mater.}, 2:539, 1990.

\bibitem{WisZenRedMon94}
E.~{Wischerhoff}, R.~{Zentel}, M.~{Redmond}, and O.~{Mondain}-{Monval}.
\newblock {\em Macromol. Chem. Phys.}, 195:1593, 1994.

\bibitem{NacRatBarKel95}
J.~{Naciri}, B.~R. {Ratna}, S.~{Baral}-{Tosh}, P.~{Keller}, and
  R.~{Shashidhar}.
\newblock {\em Macromol.}, 28:5274, 1995.

\bibitem{SveHelSkaAnd98}
M.~{Svensson}, B.~{Helgee}, K.~{Skarp}, and G.~{Andersson}.
\newblock {\em J. Mater. Chem.}, 8:353, 1998.

\bibitem{MacKenWarHep98}
R.~{Macdonald}, F.~{Kentischer}, P.~{Warnik}, and G.~{Heppke}.
\newblock {\em Phys. Rev. Lett.}, 81(20):4408, 1998.

\bibitem{HermannDavid97}
David~Sparre {Hermann}.
\newblock {\em Interaction of light with liquid crystals}.
\newblock PhD thesis, {G\"oteborg} {University} and {Chalmers} {University} of
  {Technology}, 1997.

\bibitem{TroOrrSahGed96}
M.~{Trolls{\aa}s}, C.~{Orrenius}, F.~{Salh{\'e}n}, U.~W. {Gedde}, T.~{Norin},
  A.~{Hult}, D.~{Hermann}, P.~{Rudquist}, L.~{Komitov}, S.~T. {Lagerwall}, and
  J.~{Lindstr\"om}.
\newblock {\em J. Am. Chem. Soc.}, 118:8542, 1996.

\bibitem{HulSahTroLag96}
A.~{Hult}, F.~{Sahl\'en}, M.~{Trolls{\aa}s}, S.~T. {Lagerwall}, D.~S.
  {Hermann}, L.~{Komitov}, P.~{Rudquist}, and B.~{Stebler}.
\newblock {\em Liq. Crys.}, 20(1):23, 1996.

\bibitem{TroSahGedHul96}
M.~{Trolls{\aa}s}, F.~{Sahl\'en}, U.~W. {Gedde}, A.~{Hult}, D.~{Hermann},
  P.~{Rudquist}, L.~{Komitov}, S.~T. {Lagerwall}, and B.~{Stebler}.
\newblock {\em Macromol.}, 29(7):2590, 1996.

\bibitem{LinHerOrtArn98}
M.~{Lindgren}, D.~S. {Hermann}, J.~{\"Ortegen}, P.-O. {Arntzen}, U.~W. {Gedde},
  A.~{Hult}, L.~{Komitov}, S.~T. {Lagerwall}, P.~{Rudquist}, B.~{Stebler},
  F.~{Sahl\'en}, and M.~{Trolls{\aa}s}.
\newblock {\em J. Opt. Soc. Am. B}, 15(2):914, 1998.

\bibitem{HerRudLagKom98}
D.~S. {Hermann}, P.~{Rudquist}, S.~T. {Lagerwall}, L.~{Komitov}, B.~{Stebler},
  M.~{Lindgren}, M.~{Trolls{\aa}s}, F.~{Sahl\'en}, A.~{Hult}, U.~W. {Gedde},
  C.~{Orrenius}, and T.~{Norin}.
\newblock {\em Liq. Crys.}, 24(2):295, 1998.

\bibitem{JerKur70}
J.~{Jerphagnon} and S.~K. {Kurtz}.
\newblock M.
\newblock {\em J. Appl. Phys.}, 41(4):1667, 1970.

\bibitem{Weber}
M.~J. {Weber}.
\newblock {\em CRC handbook of laser science and technology}, volume III:
  Optical materials.
\newblock CRC Press Inc., Florida, 1986.

\bibitem{Hunsperger}
R.~G. {Hunsperger}.
\newblock {\em Integrated optics: theory and technology}.
\newblock Springer-Verlag, third edition, 1991.
\newblock Chapter 5.

\bibitem{Tie71}
P.~K. {Tien}.
\newblock {\em Appl. Opt.}, 10(11):2395, 1971.

\bibitem{note2}
Both monomers are amphiphilic, as they form monolayers at the air/water
  interface which can be deposited onto solid subtrates with the
  Langmuir-Blodgett technique.

\end{thebibliography}
\end{document}